# Using Mimicry to Learn about Mental Representations

Greg Kochanski, University of Oxford Phonetics Laboratory, Oxford UK

Running Head: Using Mimicry to learn about Phonology
Correspondence to: Greg Kochanski, **greg.kochanski@phon.ox.ac.uk**, Phonetics, 41 Wellington Square, Oxford OX1 2JF, UK

**ACKNOWLEDGMENTS**

The beginnings of this work were initially funded by the Oxford University Research Development fund and in its middle stages by the UK's Economic and Social Research Council under grant RES-000-23-1094. Both were greatly appreciated. I also thank 'Oiwi Parker-Jones and Elanor Payne for asking awkward questions that improved this work.

**ABSTRACT**

Phonology typically describes speech in terms of discrete signs like features. The field of intonational phonology uses discrete accents to describe intonation and prosody. But, are such representations useful? The results of mimicry experiments indicate that discrete signs are not a useful representation of the shape of intonation contours. Human behaviour seems to be better represented by a attractors where memory retains substantial fine detail about an utterance. There is no evidence that discrete abstract representations that might be formed that have an effect on the speech that is subsequently produced. This paper also discusses conditions under which a discrete phonology can arise from an attractor model and why – for intonation – attractors can be inferred without the implying a discrete phonology.

# INTRODUCTION

We sometimes think of the units of intonational phonology as discrete entities: accents which fall into just a few categories (Gussenhoven 1999; Ladd 1996; Beckman and Ayers Elam 1997). In this view, accents in intonation are equivalent to to phonemes in segmental phonology (except that they cover a larger interval). They have a rough correspondence to the acoustic properties of the relevant region and accents form a small set of atomic objects that do not have meaning individually but that can be combined to form larger objects that carry meaning. For segmental phonology, the larger objects are words; for intonation, the larger objects are tunes over a phrase.

However, the analogy is not strong, and there are many differences. For instance, there is no known useful mapping from intonation phonology to meaning. (Pierrehumbert & Hirschberg 1990 point out some of the difficulties.) For words, this is accomplished by dictionaries and internet search engines. These technologies have no intonational equivalents. To date, attempts to connect between intonation and fundamental frequency contours have not escaped from academia: the results to date are either probabilistic (Grabe, Kochanski & Coleman 2005), have been theoretical and primarily based on intuition, or have been conducted in tightly controlled laboratory conditions (Ladd & Morton 1997; Gussenhoven & Rietveld 1997).

Likewise, there is no known, reliable mapping between sound and intonational phonology. Probability distributions overlap (Grabe, Kochanski & Coleman 2007) and automated systems for recognizing intonation have not become commercially useful. In contrast, the connection between acoustics and segmental phonology is made by speech synthesis and recognition systems. The mapping between sound and segmental phonology is complicated, but it is reasonably well understood, and reliable enough to be commercially useful. As a further contrast, transcription of intonation seems qualitatively different from transcription of segmental information. Intonational transcription (e.g. Grice et al 1996; Jun et al 2000; Yoon et al 2004) is far more error-prone and slower than transcription of words, even after extensive training. Yoon et al 2004 found an agreement of circa 85% between transcribers (depending on exactly what was being compared), but it is notable that at each point in the transcription, the transcribers had a choice between (typically) just two symbols. In a typical phonemic or orthographic transcription, the transcriber would attain comparable or higher precision while choosing between (about) 40 phones, or amongst thousands of possible words for each symbol.

So, in light of these differences, it is reasonable to ask whether intonation can be usefully described by a conventional discrete phonology or not. If it can be, what are the properties of the objects upon which the phonological rules operate? This paper lays out empirically-based answers to those questions and describes an experimental technique that can provide a reasonably direct exploration of the properties of phonological objects.

**Modelling Mimicry**

Figure 1 shows a simple model of speech mimicry. It is treated as a completely normal speech process: a person hears speech, perceives it, and generates a memory representation for it. Later, the person produces speech from the memory.

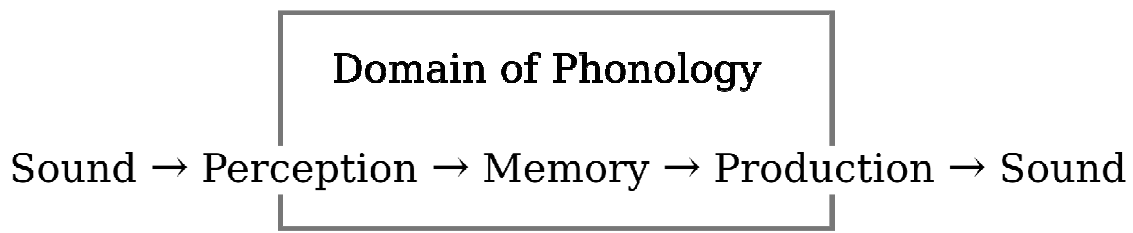

*Figure 1: Mimicry via phonology. Sound is perceived, stored in memory in some form of phonological representation, then when the subject begins to speak, he or she articulates based on the stored memory representation.*

The most contentious point might be the identification of the memory representation with a phonological representation. But, if we cannot usefully predict the acoustic properties of speech from phonology, how can phonology claim to be part of the study of language? Likewise, if phonological entities are not the end-product of the perceptual process, where do they come from?[1] This interpretation: asserts that there is some isomorphism between phonology, the mind, and the activity of the brain. In other words, that phonology can describe (at least in an approximate, abstract way) what is happening in the mind and the brain.

---

[1] Of course, the perceptual process can be described at various levels of detail, and phonology is only one level of description. However, for phonology to be meaningful, there must be a consistent description of the perceptual process that takes acoustics on one side and yields phonological entities on the other, because that is what humans do, and we would like to describe human language processing.

Some linguists would deny this biological connection, claiming that phonology is strictly a human invention that allows us to conveniently represent speech patterns in a way that humans can easily interpret and study.  But, the denial does not follow from the invention: the self-evident fact that phonology is a human invention does not prohibit it from being isomorphic to processes in the brain.  For example, secondary-school models of atoms are human constructs and some ideas of basic chemistry, such as "valence" are as abstract as phonology, but they describe – in an approximate way – the quantum mechanical behaviour of the real underlying atoms.  Thus, linguists that deny the biological relevance of phonology are not doing it out of necessity, but rather, they are making it an axiom of the field, based on tradition, history, and convenience.

Such a denial is a choice, simply reflecting the researcher's view of where to set the academic boundary. Should the field be determined by the behaviour it explains or by the representations that it uses?  Here, the intent is to study linguistic behaviour of objects simpler than words, using whatever representation is most appropriate.  The goal is to find the representation that best describes human behaviour, chosen from amongst those representations that might fit into the rest of linguistics.[2] One might

---

[2] The opposite viewpoint would be to separate the object and description, then simply accept the possibility that the description is based upon discrete categories while the object of study might be continuous.  Certainly, one can operate this way in a self-consistent manner, but there is a cost: it

reasonably hope that those representations which give the best description would have some analogy to the structure of the mind and/or brain.

Models of mimicry other than Figure 1 are possible, but they lead to a more complex description of human behaviour. For instance, if mimicry were a hard-wired part of early language learning, one might imagine that there were two separate parallel channels, one for non-phonological mimicry and one for speech that is treated phonologically. However, such a model would be more complex and evidence for a separate channel is weak[3].

Assuming the model in Figure 1, the central question is then the nature of the memory trace. Is it continuous in the sense that a small change in fundamental frequency always corresponds to a small change in the memory representation? This would imply that memory stores something analogous to pitch, suggesting a version of the Exemplar model (e.g. Goldinger 1992, Johnson & Mullenix 1997, Pierrehumbert 2001). Or, alternatively, is the memory representation discrete, leading to a model close to Generative Phonology (e.g. Liberman 1970). These two hypotheses will be considered below.

---

becomes harder to distinguish good theories from bad by the process of prediction and experimental test. I would argue that for Phonology to define itself by the representations it uses (e.g. to freeze the field onto current phonological representations) would be analogous to Astronomy defining itself to exclude spectroscopy or Electrical Engineering defining itself via the gold-leaf electroscope. Should a field define itself by its tools, it will wither when important phenomena cannot be studied with those tools.

[3] Most arguments for a separate mimicry channel assume that the phonological units are strictly discrete. Under that assumption, any early learning of speech before phonology is well-established would demand a specialised mimicry channel. However, in this paper, we are asking whether intonational phonology is discrete, so this assumption begs the question.

Here, I follow common practice (see discussion in Kochanski 2006, §2) and approximate intonation by measurements of speech fundamental frequency. This approximation is undoubtedly imperfect: for instance loudness and duration are important in defining accent locations (Kochanski and Orphanidou 2008; Kochanski 2006 and references therein). While I discuss continuous vs. discrete phonologies in terms of fundamental frequency, similar arguments could be made with respect to other acoustic properties. The two alternatives for phonology are cast as hypotheses to be tested and (potentially) rejected.

*Hypothesis 1: The memory store is a continuous representation of fundamental frequency.*

In this hypothesis (schematized in Figure 2) nearby values of speech fundamental frequency in the input utterance are represented by nearby memory representations. Further, nearby memory representations yield nearby fundamental frequencies in the speech that is eventually produced. In other words, there is a continuous mapping between input fundamental frequency and the memory representation, a continuous memory representation, and a continuous mapping on the output.

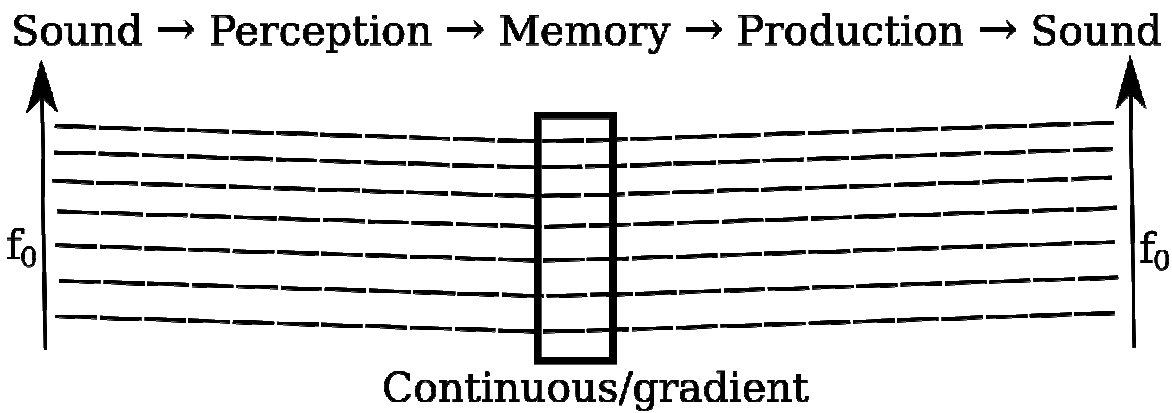

*Figure 2: Hypothetical model of mimicry where the memory store is continuous. The lower half of the drawing represents speech fundamental frequency (increasing upwards) at some point in a phrase. The lines connect input fundamental frequency (left axis) to the corresponding memory representation (centre) to the fundamental frequency that is eventually produced (right axis).*

Absent variability, the output would perfectly preserve any distinctions that were made in the input. This is not to say that the output would necessarily equal the input, though. For instance, the human who is doing the mimicry might transpose all frequencies down to a more comfortable level, as in Figure 3.

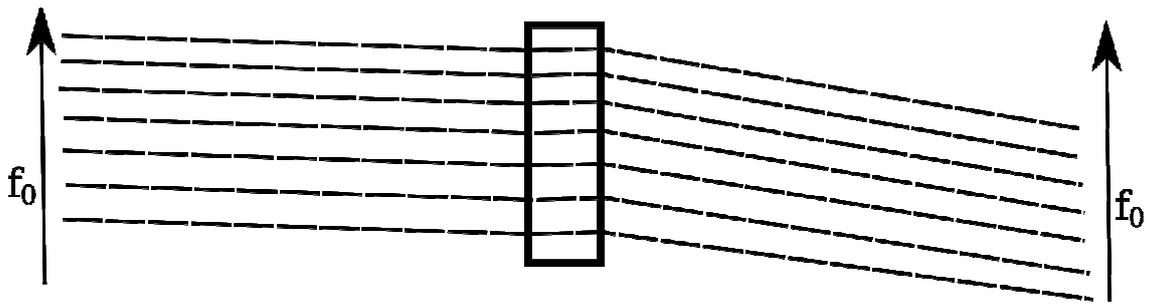

*Figure 3: Continuous mappings and memory representation for a person who is transposing down to a lower pitch. Compare with Figure 2.*

Utterance-to-utterance variation will limit the number and subtlety of distinctions that can be preserved by the mimicry process. Figure 4 shows this effect. In this example, any distinction between adjacent (e.g. the bottom two) frequencies is lost. This is a real effect in language that tends to prevent subtle phonetic distinctions from being used to represent any important phonological differences. Distinctions that are smaller than utterance-to-utterance variation will frequently be lost, leading to miscommunication and confusion. Presumably the language would evolve to avoid such unreliable distinctions.

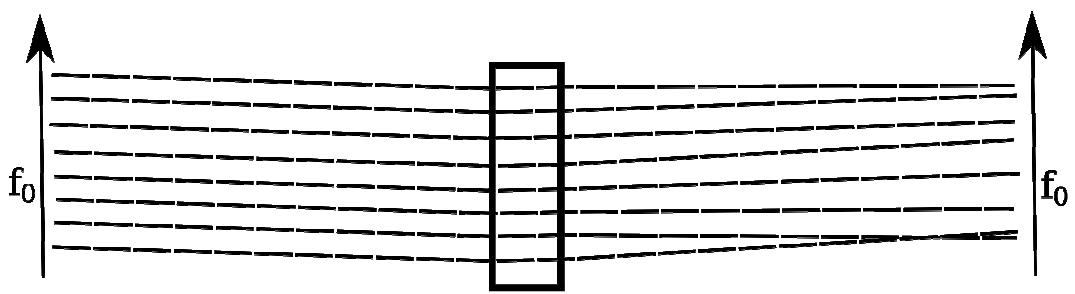

*Figure 4: Mimicry with variation in production.*

However, while language users are limited by variation, laboratory experiments need not be. Experiments can average over many utterances (a luxury that language users do not have in the midst of a conversation), reducing the variation as much as needed. If we do so, we can construct an ideal variation-free model which would look much like Figure 2. In that averaged model, all input distinctions are preserved through the memory representation to the output, even if they are not always preserved in each individual utterance (Figure 4).

Hypothesis 0: The memory store is discrete. Intonational Phonology, like most of linguistics, assumes that its object of study can be represented well by discrete symbols. For the sake of argument, we assume that we can find a minimal pair of intonation contours that differ only by a single symbol, e.g. **H** vs. **L**[4]. Figure 5 shows this hypothesis schematically. Under the null hypothesis, the intonation is perceived (either categorically or not), then stored in memory as one or the other of two discrete representations. Finally, when the subject mimics the intonation contour, his/her speech is produced from the memory representation.

---

[4] However, the argument presented here does not depend upon having a minimal pair or upon having a simple difference. We will merely assume that there are a finite number of discrete memory representations. We also assume that these memory representations are not so numerous that perception is ambiguous.

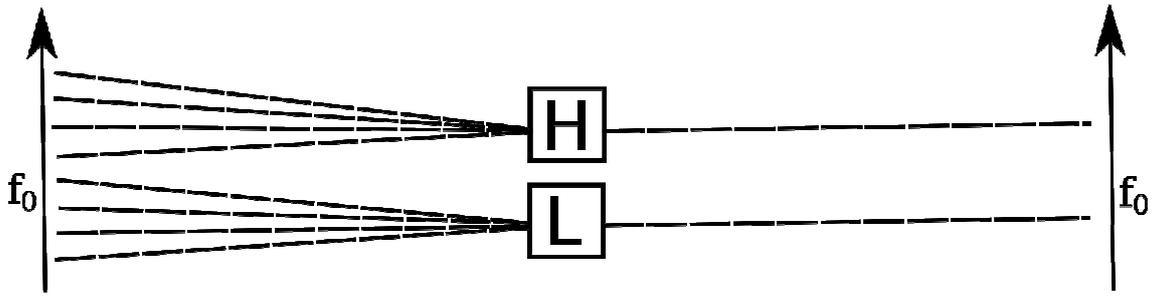

*Figure 5: Hypothetical model of mimicry where the memory store is discrete. The drawing represents speech fundamental frequency (increasing upwards) at some point in a phrase. The lines connect input fundamental frequency (left axis) to the corresponding memory representation (centre) to the fundamental frequency that is eventually produced (right axis).*

Now, production variation will yield a broad range of outputs for each memory representation. Figure 6 shows several potential outputs from the same phonology (each one corresponding to an instance of the same utterance). Potentially, the resulting probability distributions produced from **H** and **L** could even overlap (though any substantial overlap would mean that the **H** vs. **L** distinction was not sufficiently clear to form a minimal pair).

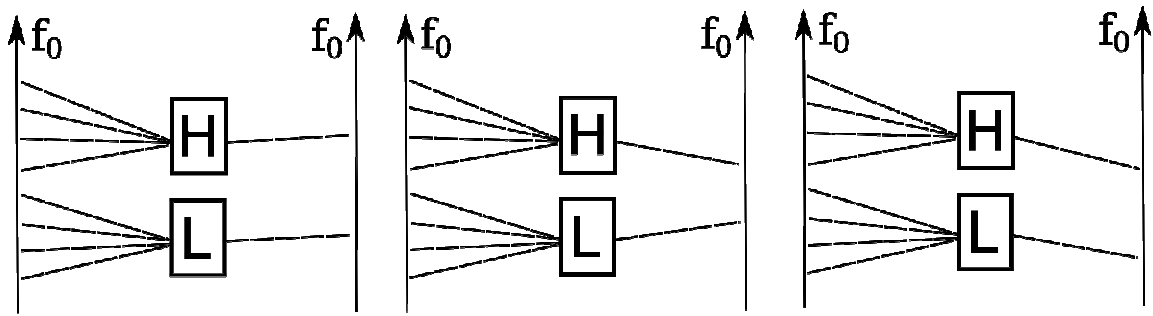

*Figure 6: Hypothetical model of mimicry where the memory store is discrete. The drawing represents speech fundamental frequency (increasing upwards) at some point in a phrase. The lines connect input fundamental frequency (left axis) to the corresponding memory representation (centre) to the fundamental frequency that is eventually produced (right axis).*

However, just as with Hypothesis 1, we can average over all productions from the same phonology and remove the effect of the variation. In this case, we see that the averaged productions form two well-separated values, different for **H** and **L**. However, the crucial difference between Hypotheses 0 and 1 lies in which distinctions are preserved. Hypothesis 1 preserves all input distinctions through to the output, but that is not the case for Hypothesis 0.

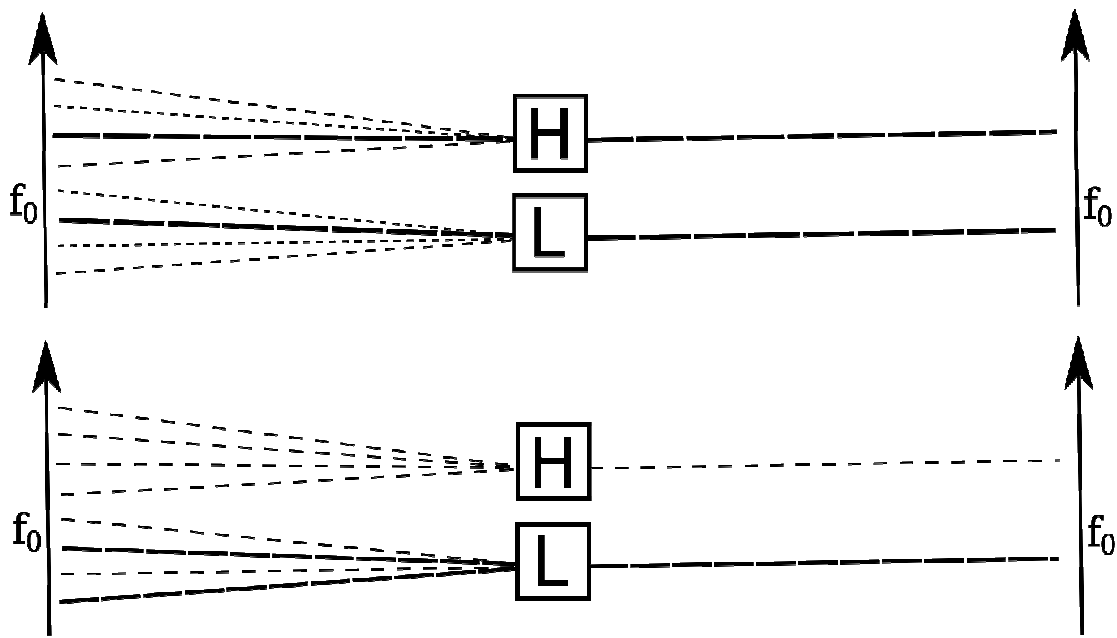

*Figure 7: If the memory representation is discrete, then only some input distinctions are preserved into the subject's mimicry output. The distinctions that are preserved are those that change the memory representation from one phonological entity to another. In the figure, the coloured lines show a pair of input stimuli (left). In the upper subfigure, the input distinction is preserved to the output because one activates **H** and the other activates **L**. In the lower sub-figure, both possible inputs (coloured/grey) lead to the same memory state, so the outputs of both coloured inputs will be identical, produced from **L**.*

Figure 7 shows that distinctions between phonological entities are preserved but not input distinctions that produce the same phonological entity. In other words, any inputs that yield the same memory representation will produce the same output, so distinctions within those sets are lost.

This behaviour is a general property of many systems and can be derived from Information Theory (e.g. Gray and Neuhoff 2000 and references therein) as discussed in (Kochanski 2006). It can be summarized as follows: *the memory representation must be complex enough to store all distinctions that are preserved to the output.* Information Theory is well established and is the ultimate basis for all modern communication technology; so, this result can be derived with mathematical rigour, though one needs to be careful about the definitions involved[5].

*Summary of Hypotheses*

The two hypotheses yield different predictions about which input distinctions people will be able to mimic reliably. This is exactly what is wanted because it will allow us to disprove one or the other hypothesis.

Equally important, we have a general principle that the memory representation must be able to store all the distinctions that people can mimic. This gives us a way to set a lower limit to the complexity of the memory representation of intonation based on observations of human behaviour. This allows us to experimentally measure at least one property of phonological entities.

**Experiments on the Intonation of Speech**

The main experiment discussed in this work have been reported in (Braun et al 2006). The goal of this paper is not to present that work again, but rather to interpret it in the light of Hypotheses 0 and 1 to see what can be learned about human memory for intonation.

---

[5] Information theory is normally applied to long messages where each can be interpreted in isolation. Applying it to human speech implies that one must consider a "message" to be a sequence of speech that is long enough so that any context outside the sequence is relatively unimportant. In practise, this means that messages should be at least a sentence long (and possibly much longer under some circumstances). Specifically, it should be noted that the figures are schematic, and should not be interpreted to suggest that individual fundamental frequency values form a valid message from the viewpoint of Information Theory.

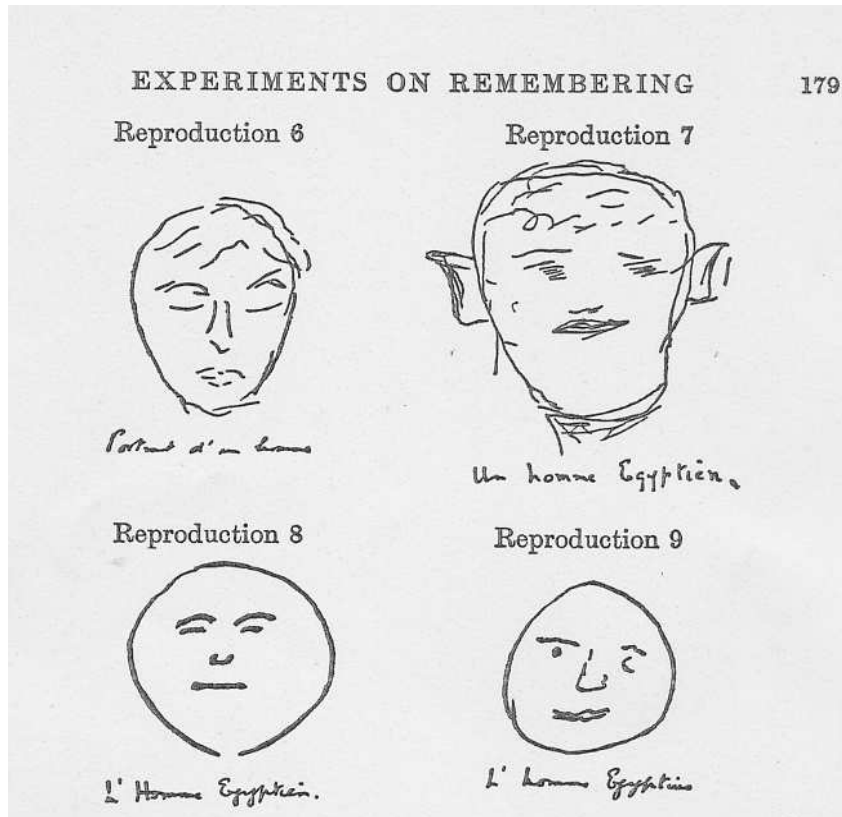

*Figure 8: Bartlett's experiments on memory and mimicry of drawings. One of the more common changes was simplification. Continued simplification of a face could potentially lead to something like the modern "Smiley." (Reproduced courtesy of Cambridge University Press.)*

The Braun et al experiment was inspired by Bartlett 1932, Pierrehumbert & Steele 1989, and Repp & Williams 1987. Bartlett conducted a mimicry experiment on images, with a group of subjects. The first subject would be (briefly) shown a drawing, and then would be asked to sketch it. In turn, that drawing would be briefly shown to the next subject, et cetera. Bartlett found a variety of changes in the drawings, but one of the more common changes was simplification (Figure 8). If one extrapolates the simplifications forward, one might well obtain something like the modern smiley, a maximally abstract representation of the human face.

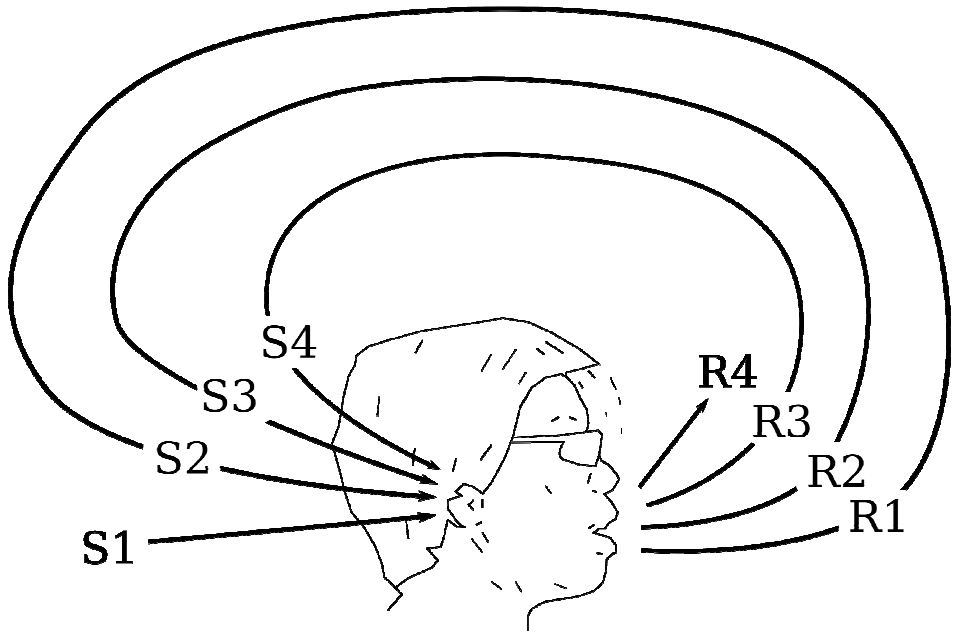

*Figure 9: The general plan of the Braun et al mimicry experiment. Subjects were asked to imitate the speech and melody of each sentence, but to use their own voice. The first stimulus, S1, was synthesized to match the subject's normal pitch range. Further stimuli (S2, ...) were the subject's own responses (after mild processing).*

The Braun et al experiment studied intonation contours rather than drawings, and it simplified the experiment by using only a single subject for each sequence. (The experiment ran in blocks of 100 utterances, presented in random order, so that the subjectd would not be able to follow an utterance from iteration to iteration.) Figure 10 shows a schematic of the stimulus flow.

Following an utterance from one iteration of the Braun et al experiment to the next, one sees a combination of utterance-to-utterance variation and systematic change from one iteration to the next. A sample is shown in Figure 11. The question arises then, is this a secular decrease or does it have a target? A secular decrease might imply nothing more interesting than imperfect mimicry in that the subject has a tendency to produce speech with a frequency slightly lower than whatever he or she hears.

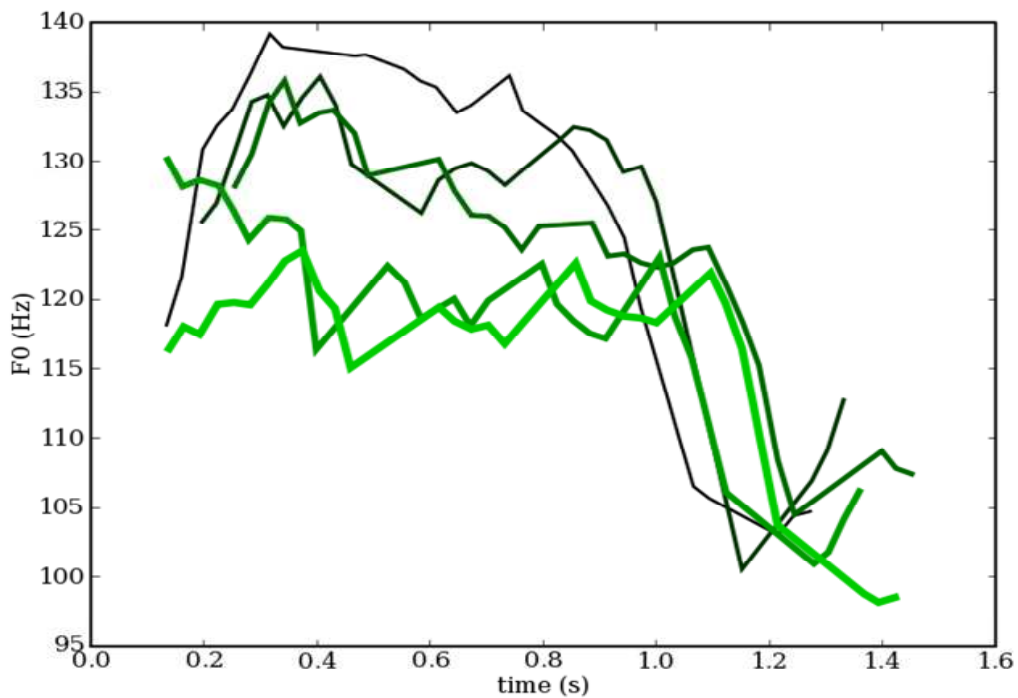

*Figure 4: Stimulus 1, then Responses 1 ... 4 of the Braun et al mimicry experiment (dark and narrow → grey and broad lines, respectively). The horizontal axis is time in each utterance and the vertical axis is the fundamental frequency of the speech. At t=0.8 seconds, the utterances are in order from S1 at top down to R4 at bottom. In the central, relatively flat region, there is a systematic decrease in fundamental frequency.*

The question can be answered by plotting the combined distribution of frequency measurements from all utterances and watching the distribution change from iteration to iteration. A downward shift would simply cause the histogram to move downward from one iteration to another. Instead, the histogram gradually became narrower and multimodal. Figures 12-15 show the intonation of a block of 100 utterances changing over four iterations. Figure 12 shows the stimuli (S1) which are linear combinations of three normal intonation contours. The feature to notice in the plot is that near the middle of the utterance (for $\tau$ between 0.3 and 0.6) the distribution of frequency measurements is broad and flat: in the stimuli, all frequencies are roughly equally probable.

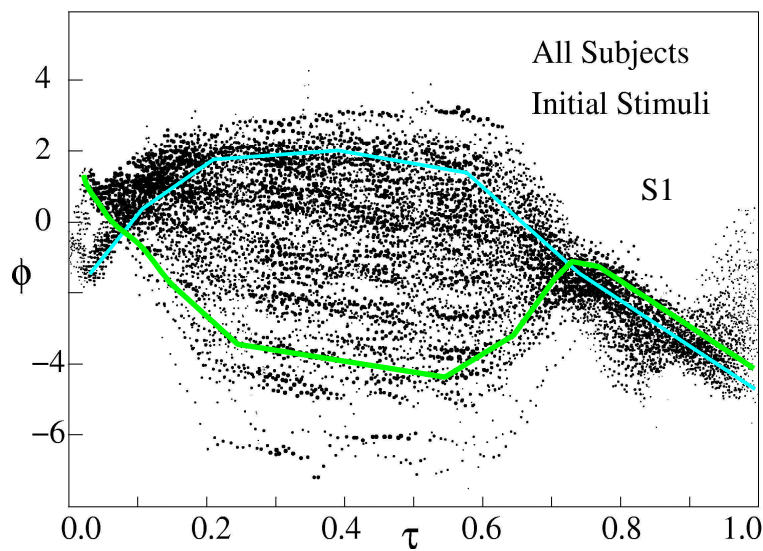

*Figure 12: The distribution of all initial stimuli. Data from one hundred utterances are superimposed to make the plot. Each dot corresponds to one fundamental frequency measurement from one utterance. The coloured lines trace out two of the 100 utterances. The horizontal axis ($\tau$) is normalized time and the vertical axis ($\varphi$) is frequency in semitones relative to the subject's average frequency.*

However, after just one mimicry (iteration), the situation has changed. Figure 13 shows R1/S2. The variability of the fall where τ is near 0.8 has decreased, and the upper edge in the middle of the utterance has become denser.

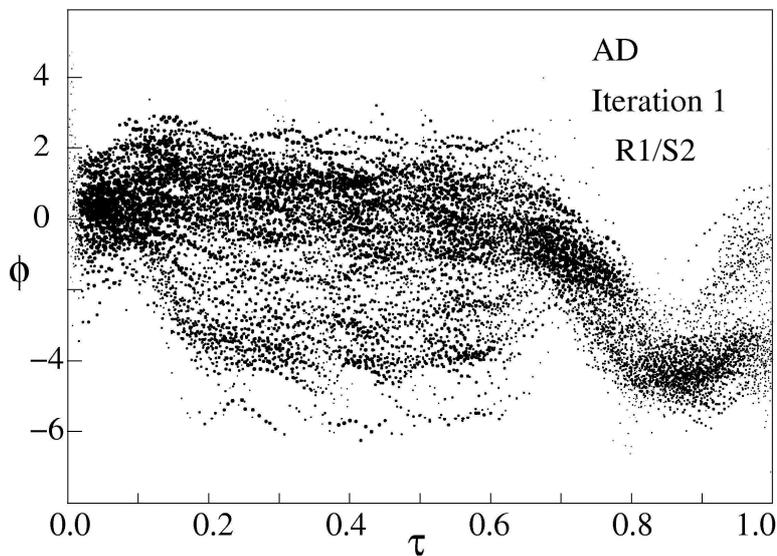

*Figure 13: scatter plot of frequency measurements for subject AD after utterances have been mimicked once. Plotted as per Figure 12.*

After a second mimicry (Figure 14), the upper edge, near the middle of the utterance is becoming a density peak about 1 semitone above the speaker's average frequency, and another clump is forming, about three semitones below the speaker's average frequency. Another effect is that relatively few samples are found in between the clumps: the region where τ is near 0.25, one to two semitones below the speaker's average, is becoming sparse.

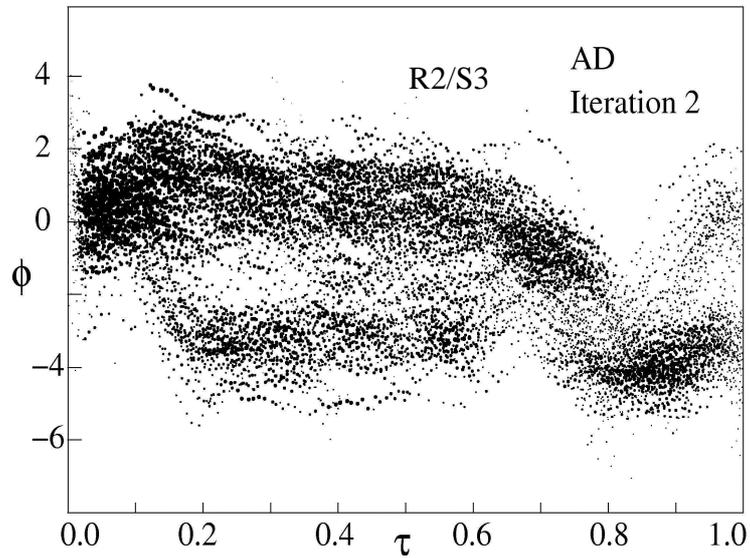

*Figure 14: scatter plot of fundamental frequency measurements after two mimicries. Plotted as per Figure 12.*

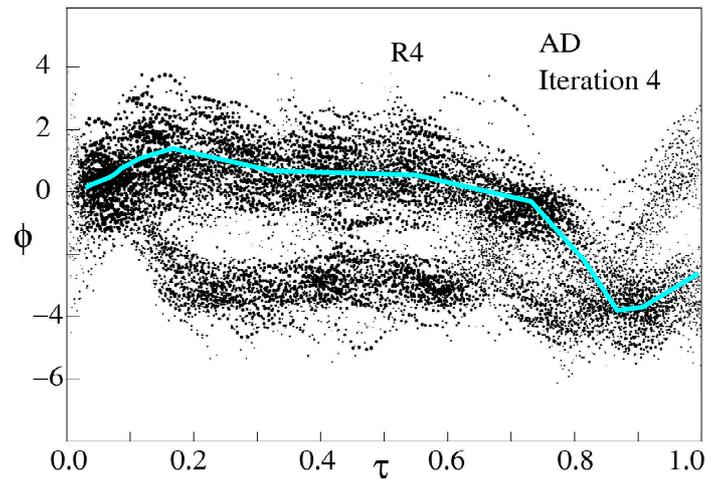

*Figure 15: The scatterplot at the end of the experiment, after four mimicries. Plotted as per Figure 12. The blue line marks one utterance's intonation contour.*

Finally, after four mimicries, Figure 15 shows that two separate groups of intonation contours have formed in the central part of the utterance. Utterances with intermediate frequencies have almost disappeared.

What is happening is that every time an utterance is mimicked, the produced intonation contour is biased towards one or the other of these two groups of contours. Figure 16 shows this by comparing an early and a late production. Aside from a certain amount of random variation, the contours approach either a high target or a low target, whichever they are closest to. In mathematical terms, from one iteration to the next, the contours are mapped towards one of these two attractors.

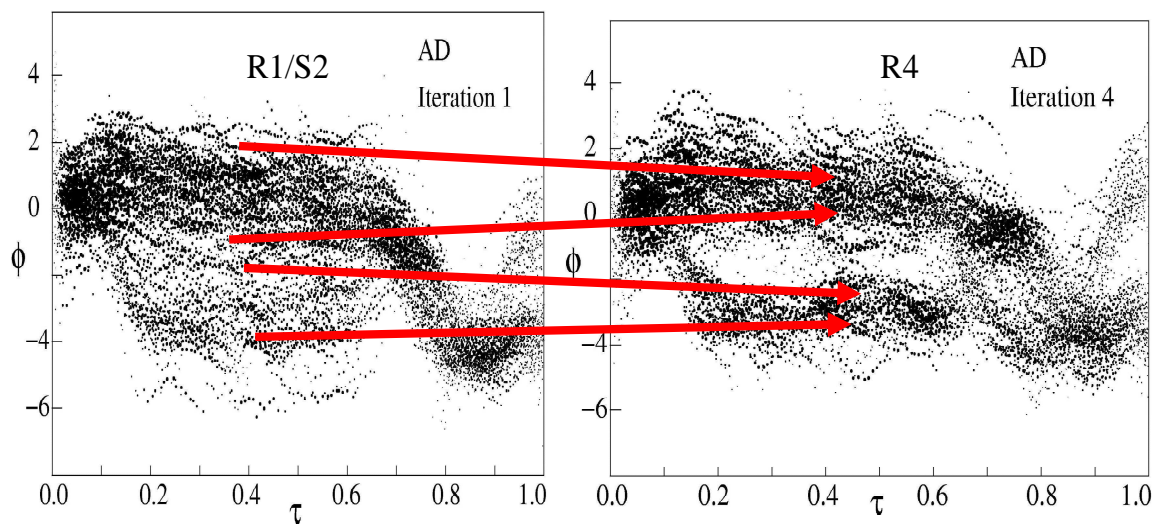

*Figure 16: changes in the scatter plot between early and later productions in the mimicry experiment. From iteration to iteration, contours follow the red arrows: the highest stimuli are mimicked lower, the lowest are mimicked higher, and contours in the middle move up or down, depending on whether they are closer to the high group or the low group.*

*An Engineering Analogy*

There is a close engineering analogy to this behaviour. It is called the "Static Discipline" and is taught in undergraduate electronic design classes. It is an essential design rule that makes digital electronics possible. (One might be tempted to suppose that an equivalent design rule evolved within the brain.)

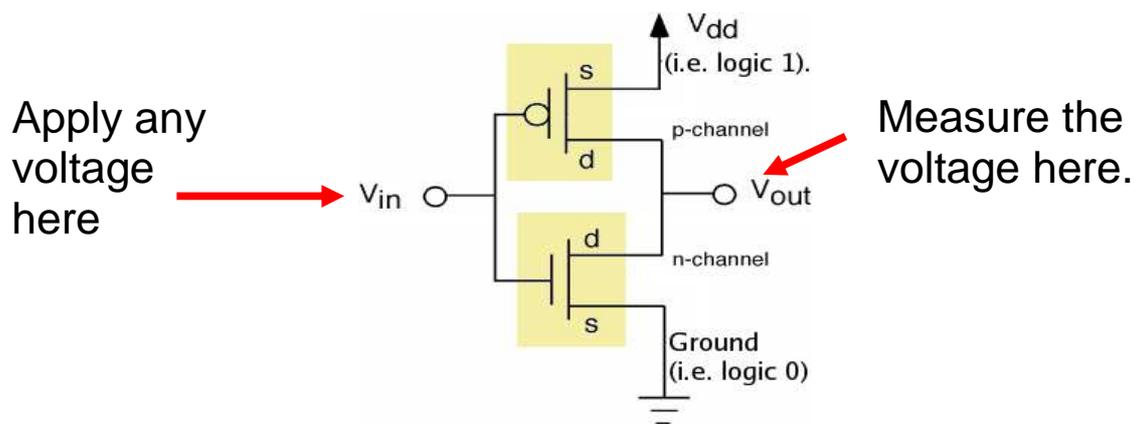

*Figure 17: C-MOS inverter circuit.*

Consider the simplest logic gate, an inverter (Figure 17). It is typically constructed out of two CMOS transistors, one N-channel and one P-channel. The two transistors have complementary properties so that when the input voltage is high, the lower transistor conducts and the upper transistor is off. As a result, the output voltage is pulled low. When the input voltage is low, the top transistor is turned on, the bottom one is turned off and the output voltage becomes high.

This device relates each input voltage to a corresponding output voltage. Mathematically, it maps between its input and its output (Figure 18). There is also a small amount of noise, which might play the same role as utterance-to-utterance variation in language. Both sub-figures display the same input-to-output mapping; they just show it in different ways.

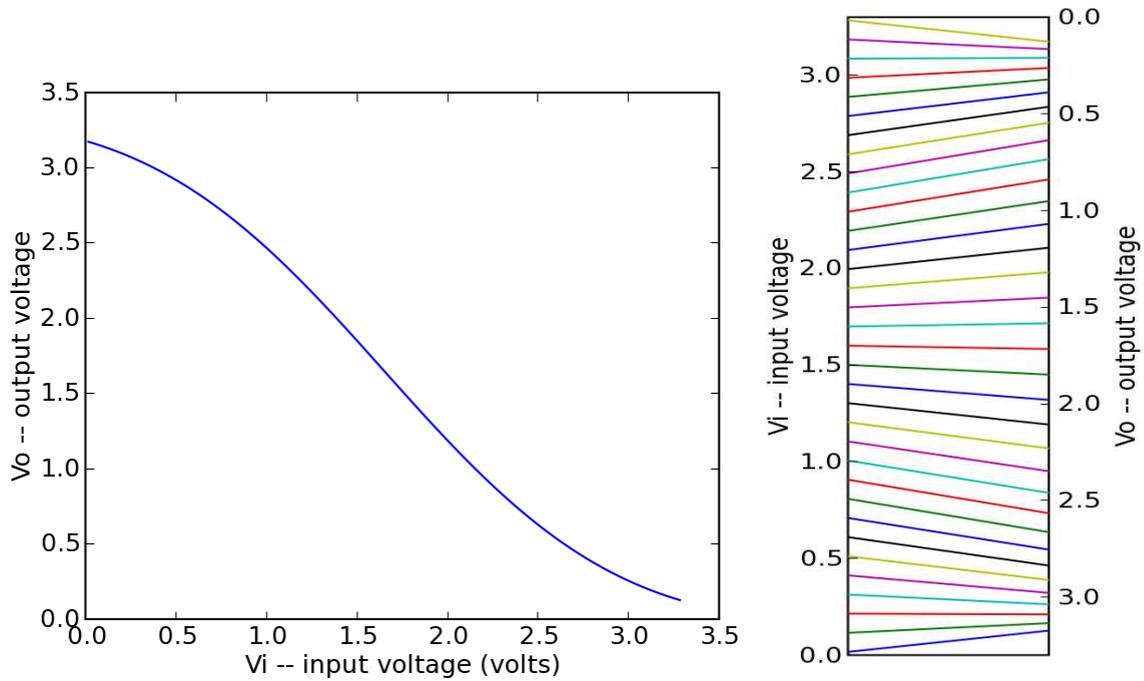

*Figure 18: The C-MOS inverter's input-to-output mapping. The input voltage is placed on the left axis, and the output voltage is on the right axis. Lines connect corresponding input/output pairs. The mapping is compressive near the top and bottom where a given range of input voltages yields a smaller range of output voltages.*

The static discipline requires that any digital logic element should have two regions where the mapping is compressive: one near zero volts input, and one at relatively high voltage. These compressive regions are important not so much in the context of a single logic gate, but rather for their effect on a large system composed of many logic gates connected in series. Computers, of course, are large and complex systems where any signal that is fed into one of the pins of a processor may propagate through at least dozens of logic gates before it comes out on some other pin. So, we can idealize a computer as a string of C-MOS inverters (Figure 19).

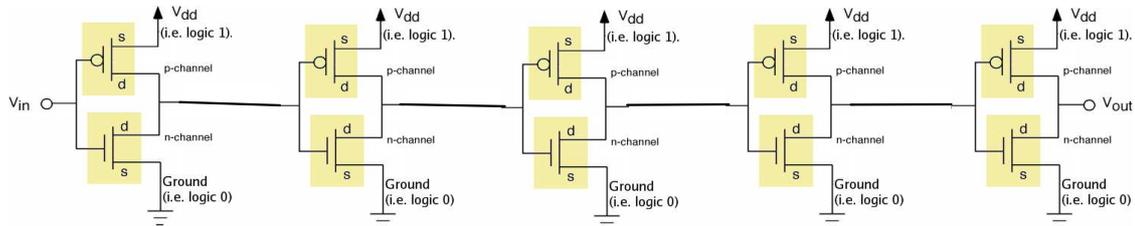

*Figure 19: A string of C-MOS inverters. We will imagine putting a voltage on the first input, then measuring the voltage at all intermediate points in addition to the final output.*

Each C-MOS inverter has a mapping from its input voltage to its output voltage. Likewise, every iteration of the Braun et al mimicry experiment reveals a mapping from the fundamental frequency of the stimulus to the fundamental frequency of the mimicked response. We can make an analogy between the two.

At this point, we have the tools needed to simulate a string of C-MOS inverters or (equivalently) a sequence of iterated intonational mimicries. The crucial ingredient is Figure 18, the mapping from input to output of each stage. One simply considers the stages (or iterations) one at a time, applying the Figure 18 mapping at each step. Since the output of one stage is the input to the next, we just take the output of the first mapping and use it as input for the second, then take the output of the second and use it as input for the third, ad infinitum. The result of this repeated mapping is shown in Figure 20. Each vertical line corresponds to the output of one inverter and the input of the next (or, by analogy) the response to one iteration of the mimicry experiment and the stimulus for the next.

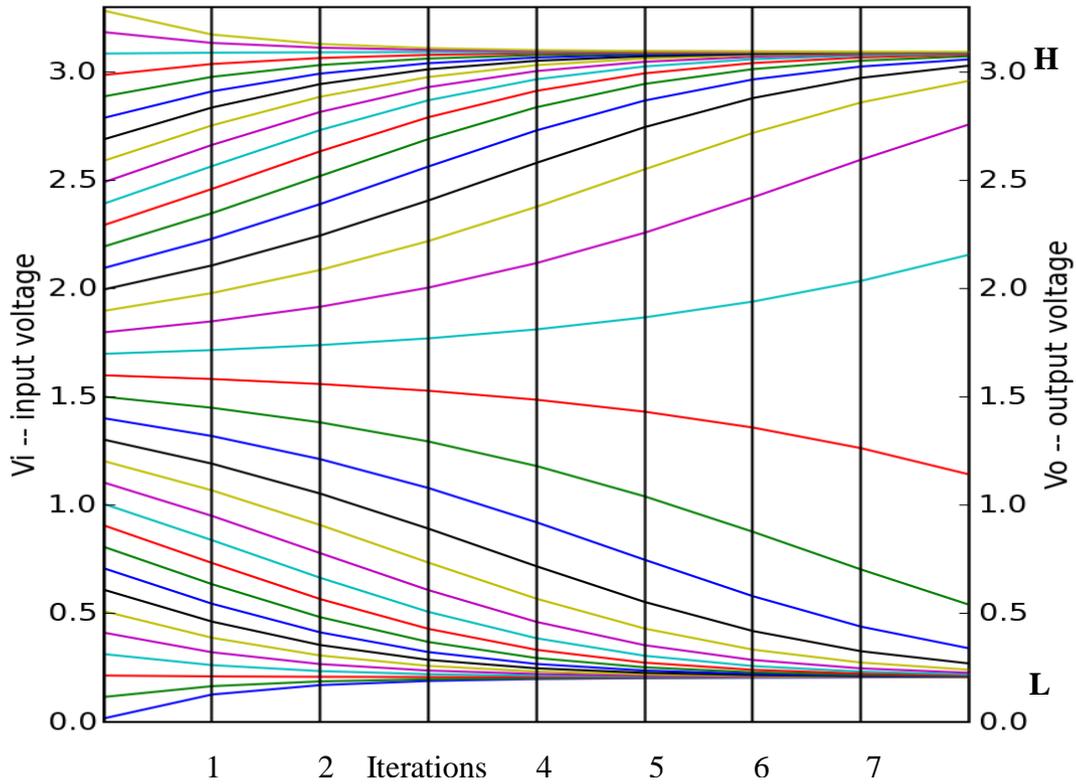

*Figure 20: voltages within a string of C-MOS inverters. The output of each inverter drives the input of the next.*

As can be seen toward the right side of Figure 20, this iterated system has an interesting behaviour: after enough stages, almost any input voltage gets mapped to either 0.2 V or 3.1 V. The system gradually becomes digital as it is made longer. This is the result of a series of compressive mappings. Each stage compresses voltages near 3.1 V together and it also compresses voltages near 0.2 V closer together. Conversely, the mapping of Figure 18 magnifies voltage differences near 1.7 V: different voltages near the mid-range get pushed further and further apart.  In the limit of an infinite string of inverters, any input would yield an output voltage that could be precisely represented as a digital **H** or **L** state. This is an example where a discrete, digital system appears as an emergent property from analogue/continuous components.

Voltages between **H** and **L** do not stay there, they move away from the centre towards either the high attractor or the low attractor, whichever is closer.  This result is analogous to what is seen experimentally in Figures 12-15, and it seems fair to interpret those figures as the result of an iterated mapping with two compressive regions.   Each compressive region, after a few iterations, yields a dense group of intonation contours.

The static discipline is a design rule, and as such it has a purpose. The purpose is to force a complex system built out of these inverters to have two attractors.  This allows the system to be thought of as digital, with discrete states.  In a system built under the static discipline, there is no way to incrementally convert a low voltage into a high voltage by small  changes because each C-MOS inverter will pull the voltage back towards the nearest attractor.  This return toward the attractors is crucial in real systems because it means that small amounts of noise will not cause errors.  Even if each stage of the system adds noise that pushes the voltage away from the attractors, the next stage will un-do the noise, pulling the voltage back towards the attractors.  It is tempting to say that this is the mechanism by which discrete phonologies emerge from a continuous/analogue brain.  It is tempting to see this as a victory for Hypothesis 0. While that might be the correct conclusion for segmental phonology or words, we will see that it is not true for intonation.

**DISCUSSION**

**Intonational Attractors are Slow**

We saw already that it took several iterations of the mimicry experiment for the intonation contours to approach the high and low attractors. This can be quantified by measuring how strongly bimodal each scatter-plot of fundamental frequency is (e.g. Figure 15). Without going into the details (which can be found in Braun et al 2006), the results can be seen in Figure 21. That figure is the answer to the question "How strongly bi-modal is the frequency distribution?" The vertical axis (valley depth) measures how empty is the middle of the scatterplot (e.g. Figure 15), relative to the density of fundamental frequency measurements near the high and low attractors. A value of zero implies that there is only a single maximum (not bimodal at all); values smaller than one imply two strongly overlapping peaks; values greater than one indicate two well-separated peaks with larger values indicating increasing separation.

 The gradual increase in valley depth from iteration to iteration implies a slow and gradual separation of the scatter-plots into two peaks, over the course of several iterations. Recall that each iteration is a complete pass through the human subject involving on the order of 100 stages where one neuron triggers another[6], so if we equate a logic gate with a few neurons, the rate of convergence per group of neurons (i.e. per logic gate) must be small indeed.

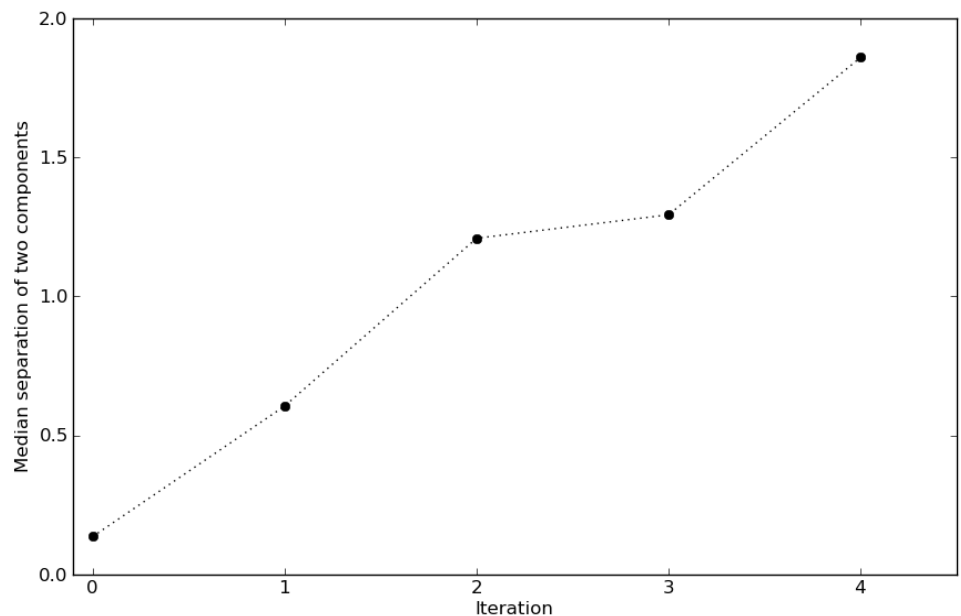

*Figure 21: A measurement of the bimodality of f0 near the centre of the utterances. The horizontal axis shows the number of experimental iterations, starting with the initial stimulus. The vertical axis displays the separation of the two (presumably phonologically distinct) components of the fundamental frequency distribution.*

More practically, if it takes roughly four iterations for the fundamental frequency to converge toward a pair of almost-discrete states, then one certainly should not expect digital behaviour to emerge on a single trip between the ears and memory. The convergence that we see is approximately ten times too slow for intonational phonology to be accurately represented by a discrete memory representation.

---

[6] A typical interval between neuron firings is about 10 milliseconds, and these intonation contours were remembered by the subjects for about 1 second. Thus, a memory of an intonation contour in the experiment is preserved across about 100 generations of neuron firings.

**What is stored in the memory representation?**

One should also consider which distinctions the subjects can mimic. Recall that the memory representation must be at least rich enough to store all the distinctions that can be mimicked. A comparison of Figures 12 and 13 shows that subjects are able to mimic fine phonetic detail fairly accurately. Not only can subjects reproduce the contours that happen to be near the attractors, but they can reproduce the extreme contours and the contours between the attractors, too. So, all this detail is stored in memory and is plausibly part of the phonological entities.

Hypothesis 1 is actually the better approximation to our data, at least over a single iteration. All input distinctions are carried through to the output, although some distinctions may be emphasized and others reduced. Figure 22 shows one reasonable interpretation for mimicry behaviour. This model takes the view that the memory representation is essentially an acoustic memory, but biased slightly toward one or another special intonation contours. If interpreted literally, this model suggests that intonation contours are stored in something like the phonological loop (Baddeley 1997) and the gentle bias toward the attractors is due to interactions with something stable outside the phonological loop.

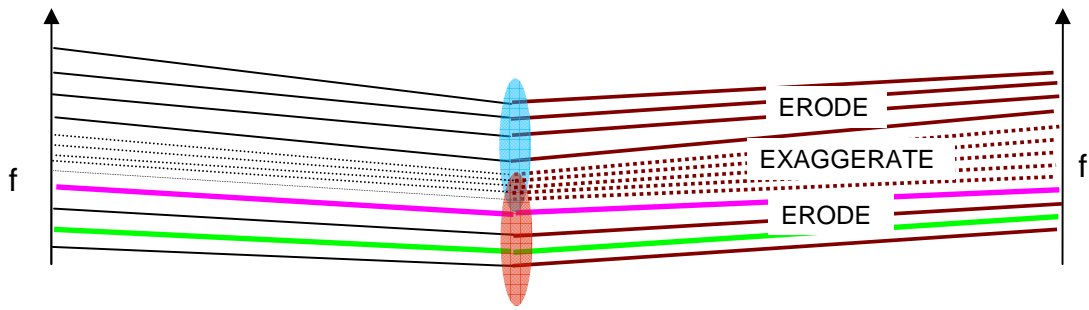

*Figure 22: A plausible interpretation of the mimicry results, corresponding to an intermediate case between Hypothesis 0 and Hypothesis 1. All distinctions are preserved, but some are partially eroded and others are emphasised.*

Another reasonable interpretation that is closer to the traditional phonological approach is to consider the memory to be a discrete phonological symbol along with substantial amounts of fine phonetic detail. This is a sort of "decorated object", shown in Figure 23. However, this interpretation does not carry a license to do traditional discrete phonology. The fine phonetic detail exists, stored in the memory representation, so one cannot arbitrarily ignore it. A proper phonological theory would include it, would involve it in the computations, and presumably, the fine phonetic detail would affect the answer generated in some phonological computations.

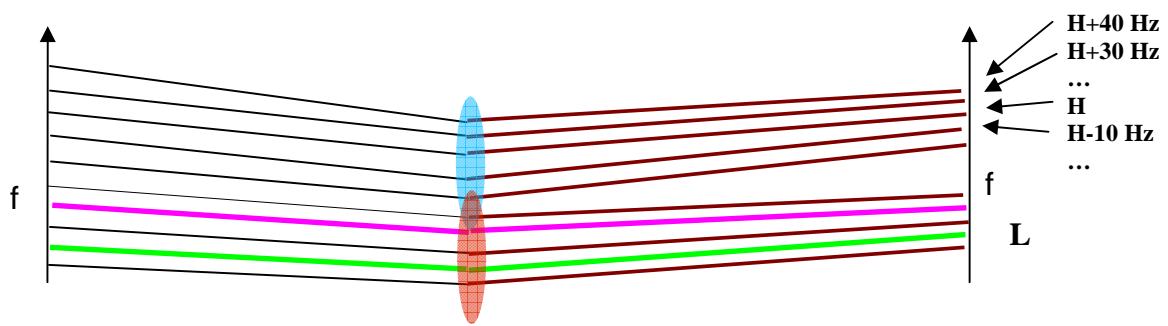

*Figure 23: A plausible interpretation of mimicry results in terms of decorated categories or decorated symbols.*

Given that some fine phonetic detail is stored, the onus is on the phonologists to show that their computations are useful descriptions to human language behaviour and that ignoring the phonetic detail is a valid approximation. Any phonological theory that uses discrete objects carries a implicit assumption that such discrete representations actually exist in the mind or at least are a good description of how the mind works. This is a strong assumption and needs to be justified, otherwise the resulting theory is built on sand.

We know the fine phonetic detail is used because we can hear the detail when a subject mimics an intonation contour. Since the detail is in the memory representation and accessible to conscious introspection, it seems likely that the phonological processes of speech production do not limit themselves to using only the discrete part of a decorated object. They use both the discrete part and the fine phonetic decoration, and presumably other phonological processes do too. The challenge is on the theorists to re-cast phonology in terms of either of these interpretations.

**CONCLUSION**

A straightforward interpretation of results from mimicry experiments shows interesting, complicated behaviour. The existence of attractors in intonation and their similarity to common intonation contours suggests that something like intonational phonology exists. However, the approach toward the attractors is far too slow for discrete phonological categories to be a good approximation to the way humans actually remember and reproduce intonation. To the extent that discrete phonological entities exist for intonation, they have only a weak influence on actual behaviour.

Humans do not behave as if their memory representation of intonation were a few discrete states. Memory certainly captures a much richer set of distinctions than two phonological categories, and a reasonable interpretation is that a substantial amount of detailed information about the intonation contour is stored in memory, available for processing. Further, this detailed information is actually used in the mental processes of speech production.